# Developments in Readout for Silicon Microstrip Sensors at a Linear Collider Detector

Jerome Carman, Kelsey Collier, Sean Crosby, Vitaliy Fadeyev, Bruce A. Schumm, and Ned Spencer

Santa Cruz Institute for Particle Physics, and Department of Physics
University of California at Santa Cruz, Santa Cruz CA 95064

We report recent results on the use of charge division to obtain a longitudinal coordinate from silicon strip detectors, and on sources of electronic readout noise for long, thin strips. These results hold promise for the design of Linear Collider charged particle tracking system composed of silicon microstrip sensors.

## 1 Resistive Charge Division

The principle of resistive charge division for position measurement was laid out by Radeka in 1974 [1] but, to date, has not been explored for estimating the longitudinal coordinate for silicon microstrip detectors. To exploit resistive charge division, the impedance of the readout structure must be large relative to the dynamic impedance of the amplifier circuitry. Somewhat counter-intuitively, [1] suggests that, at this level and above, the longitudinal resolution is independent of resistance. Thus, we have explored the use of the implant (60 k$\Omega$/cm) as a readout structure for resistive charge division.

Since no prototype sensors were available for this study, we instead developed a model of such a device by representing the continuous RC network by 10 discrete subdivisions (see Figure 1). Values of $R_n$ = 60 k$\Omega$ and $C_n$ = 1.4 pF were used to represent a 10cm long sensor for which the p$^+$ implant is read out directly, with no metal strip attached to the implant (through either DC or AC coupling).

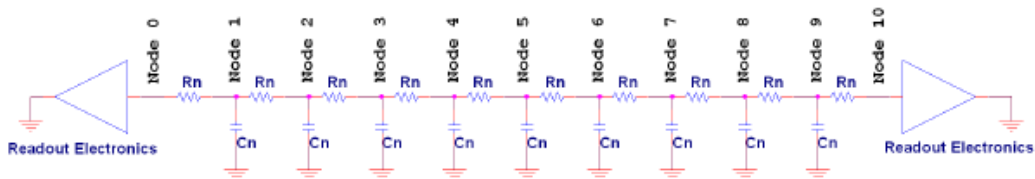

Figure 1: Discrete network used to model the distributed RC characteristics of a resistive silicon strip sensor.

The physical device represented by Figure 1 was used to benchmark a PSpice simulation, which was then extended to a network of five neighboring strips to explore the effect of capacitive coupling to neighboring channels. For the five-strip case, the 1.4 pF capacitance was split 40%/60% between ground and the two neighboring channels, respectively.

Expectations, later confirmed by the benchmarked simulation, were that readout noise would be dominated by the Johnson noise of the resistive implant. Thus, off-the-shelf (discrete) operational amplifiers were chosen to read out the physical network of Figure 1. The first stage of amplification was performed by a high-bandwidth Burr-Brown OPA657 FET input operational amplifier with low current noise of 1.3 fA/root(Hz), which was followed by one passive and two active stages of signal shaping. The active stages were instrumented with Analog Devices model ADA4851 bipolar video amplifiers; all shaping stages included pads allowing for the substitution of different-valued feedback components so that the dependence of the charge-division performance on overall shaping time could be studied.

Optimization of the shaping time was driven by two competing effects. Short shaping time led to ballistic deficit (signal arriving over a time period of order or longer than the shaping time) for pulses delivered to the far end of the network, and a corresponding degradation of the linear relation between the longitudinal position of the charge deposition and the amplified signal. Long shaping time, however, led to an increase in noise: since the strip is read out on both ends for the charge-division measurement, the effective electronic network presented by the strip contains both parallel (noise increases with shaping time) and series (noise decreases with shaping time) components. For long shaping time, the former dominates, leading to an eventual degradation of signal-to-noise as $\tau_{shape} \to \infty$. Balancing these two effects for the single-channel simulation led to the choice $\tau_{shape} = 1.8$ μs, or about 2.5 times the signal rise time, which was implemented in the readout of the network simulation board. This value is about 40% lower than that expected from [1]; this may be due, however, to a very stringent linearity spec of <0.2% required in [1], which would lead to somewhat longer shaping times at the expense of signal-to-noise. The optimum shaping time for the five-strip simulation was found to be 2.3 μs, or about 4 times the five-strip rise time.

To quantify readout noise, a signal of 3.8 fC (approximately 1x minimum ionizing) was delivered to the input of the preamplifiers, allowing the gain to be measured. Knowing the gain, noise measured through three different techniques – scope trace-merging, integration of the noise spectrum in frequency domain, and through the RMS deviation of pulses acquired through the digital storage oscilloscope – was converted to units of effective electrons. Table 1 shows that these measurements were in excellent agreement with expectations from the simulation.

| Single-Strip Model Noise Results | | |
|---|---|---|
| Measurement | Noise [mV] | Noise [fC] |
| Trace Merging | 3.67 | 0.23 |
| Spectrum Analyzer | 3.80 | 0.24 |
| Fitter Gaussian RMS | 4.01 | 0.25 |
| PSpice Simulation | 3.79 | 0.23 |

Table 1: Benchmarking comparison between expected noise (PSpice Simulation) and that of three independent measurements of the readout noise of the physical network simulation board. Good agreement is observed.

The estimated longitudinal position is given by

$$P = \frac{x}{1+x} \quad (1)$$

where $x = Q_r/Q_l$ is the ratio of the charges measured on the right and left side of the detector.

Figure 2 shows the comparison between actual and mean measured position for charge deposited at various fractional distances from the ends of the physical strip model. The dependence is quite linear, indicating that resistive charge division is the dominant effect in dividing the charge between the two ends of the strip.

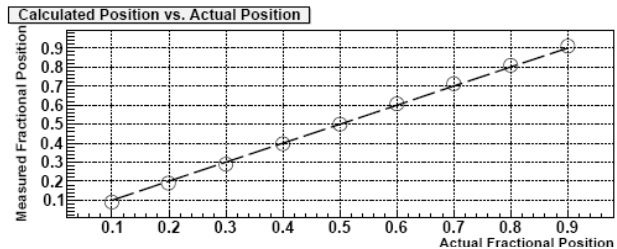

Figure 2: Measured vs. actual position for charge deposited at various fractional distances from the ends of the physical strip model.

The fractional longitudinal resolution $\sigma_f$ is given by

$$\sigma_f = \sqrt{\left(\frac{\sigma_R}{Q_R}\right)^2 + \left(\frac{\sigma_L}{Q_L}\right)^2 - \rho\left(\frac{\sigma_R}{Q_R}\right)\left(\frac{\sigma_L}{Q_L}\right)} \qquad (2)$$

where $\rho$ is the correlation between the left and right amplifier readout noise. From [1], the correlation is expected to be negative; a series of simultaneous measurements of the left and right amplifier chain outputs yielded $\rho = -0.61$. Given this, the overall noise value from Table 1, and a charge deposition of 3.8 fC, equation (2) yields a longitudinal position resolution of 6.1% of the length of the sensor, essentially independent of the position of the deposition. Simulations performed with greater (6 MΩ) and lesser (60 kΩ) total resistance, with an appropriate scaling of the electronic shaping time, exhibited the lack of dependence of longitudinal resolution on series resistance predicted by [1].

Using the five-strip simulation, it was found that the expected signal-to-noise degraded by 5%. Assuming that the correlation factor $\rho$ doesn't change, this implies a correspondingly slight degradation of the longitudinal resolution for the more realistic configuration. All together, the suggestion of a longitudinal resolution of less than 7mm for a 10cm sensor is an interesting possibility for a Linear Collider detector.

## 2  Readout Noise in the Long, Thin Strip Limit

The need to instrument a large volume with high resolution microstrips sensors, combined with a beam profile that permits long shaping-time readout, suggests that long ladders of fine-pitch sensors may be an attractive solution for a Linear Collider detector. Due to the narrowness and length of the readout strip, however, performance of a tracker composed of long ladders may be limited by the series resistance presented by the strip. According to [2], readout noise Q (in equivalent electrons) for an equivalent circuit represented by lumped resistor and capacitor elements is expected to be of the form

$$Q^2 = F_i \tau \left( 2eI_d + \frac{4kT}{R_B} + i_{na}^2 \right) + \frac{F_v C^2}{\tau} \left( 4kTR_S + e_{na}^2 \right) + 4F_v A_f C^2 \qquad (3)$$

where $R_S$ ($R_B$) is the sensor network series (parallel) resistance, C the overall strip capacitance, $\tau$ is the amplifier chain shaping time, $I_d$ the detector leakage current, $i_{na}$ ($e_{na}$) the amplifier current (voltage) noise, and $F_i$, $F_v$ are shape parameters of order unity defined in [2]. The term involving series resistance $R_S$ represents the contribution from strip resistance, which is expected to dominate in the long, narrow strip limit.

This relation was explored through the use of a single 4.75cm-long, 50 μm-pitch prototype sensor fabricated during the SiD sensor prototyping run. The traces were measured to have a resistance of 286Ω and a capacitance of 5.1pF. A strip length of up to 62 cm (3.7 kΩ and 66 pF) was measured by bonding across alternate ends of nearby strips in a back-and-forth "snake" pattern. The strips adjacent to the ones used in this study were held at ground potential.

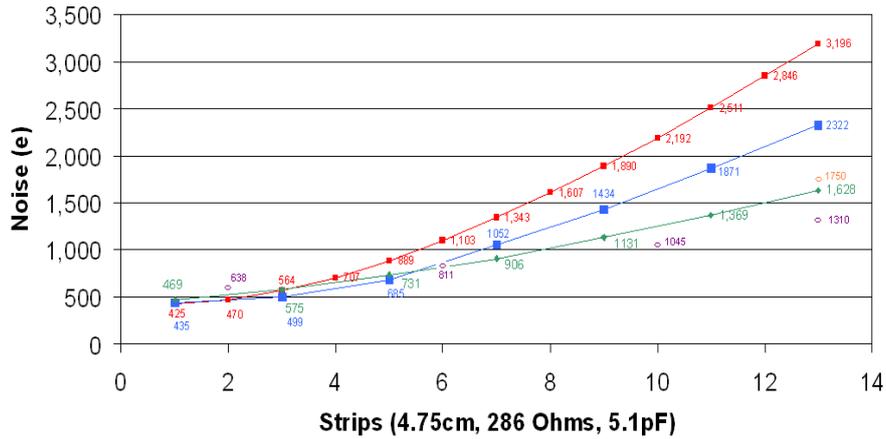

Figure 3: Expected versus observed input-referred readout noise as a function of ladder length for the "snake" sensor. The red curve represents the naïve expectation, setting the shape parameters $F_i$ and $F_v$ to 1; for the blue curve they are set to the value determined according to [2] from the shape of the amplifier excitation function. The green curve represents the experimental results. The purple points exhibit the level of readout noise observed when the sensor was read out from the center of the "snake" rather than from its end. The orange point represents a repeat of the corresponding green point after re-assessing the calibration of the injected charge (the purple points are based on the latter assessment).

Figure 3 shows the expected and measured noise as a function of the number of strips connected in series. The red trajectory is a naïve expectation based on (3) under the assumption that $F_i = F_v = 1$, which is corrected for the blue trajectory with the inclusion of values of $F_i$ and $F_v$ determined from the mean excitation profile of the amplified signal. The green trajectory shows the measured values of the noise, while the purple points represent the noise observed when the "snake" was read out from the central strip (effectively dividing the sensor network into two parallel contributions) rather than its end. The orange dot represents a repeat measurement of the 13-strip snake after a rigorous re-calibration of the measurement system.

In the regime expected to dominated by the series resistance contribution (greater than 5 or so strips), the noise is observed to be approximately 25% less than the expectation of [2]. The noise is reduced by a further 20% when the network is read out from the center. We are currently developing a model of the network and amplification chain in PSpice in an attempt to gain a deeper understanding of this result. Clearly, though, this result shows promise for the implementation of precise, long silicon microstrip ladders for a Linear Collider detector.

# 3  Summary

We have pursued two generic studies relating to the application of silicon microstrip detector to a Linear Collider detector. The first of these suggests that resistive charge division can be employed to achieve a longitudinal resolution of better than 7 mm for a 10cm-long sensor. The second suggests that readout noise may increase significantly slower with strip resistance than expectation, presumably due to distributed network effects, and that readout noise can be reduced even further if strips with significant resistance are read out from the center rather than the end. Both of these hold promise, upon further development, to yield dividends for a future Linear Collider detector.